\documentstyle[preprint,pra,aps]{revtex}
\begin{document}
\draft
\title{\bf Calculation of parity nonconservation in cesium
and possible deviation from the Standard Model}
\author{V.A. Dzuba, V.V. Flambaum, and J.S.M. Ginges}
\address{School of Physics, University of New South Wales,
Sydney 2052, Australia}
\date{\today}
\maketitle

\tightenlines

\begin{abstract}
We have calculated the $6s-7s$ parity nonconserving E1 transition
amplitude ($E_{PNC}$) in cesium.
This calculation has been performed with higher numerical accuracy than
our 1989 calculation [V.A. Dzuba, V.V. Flambaum, and O.P. Sushkov,
Phys. Lett. A {\bf 141}, 147].
Also the Breit interaction has been included and the radiative
corrections estimated.
Our final result is $E_{PNC}=0.902\Big( 1\pm (0.7\%)\Big) iea_{B}(Q_{W}/N)$.
This represents an improvement in the accuracy of the calculation
from the $1\%$ error claimed in 1989.
This result corresponds to a nuclear weak charge for Cs,
$Q_{W}=-72.39\Big( 1\pm 0.4\%({\rm exp}) \pm 0.7\%({\rm theory})\Big)$.
We conclude that there is no significant deviation from the
Standard Model value $-73.09(3)$.

\end{abstract}
\vspace{1cm}


\section{Introduction}

Measurements of parity nonconservation (PNC) in atoms provide
a very useful test of the Standard Model.
In order to extract the value of the nuclear weak charge $Q_{W}$
from experiment, atomic structure calculations must be performed.
The most precise measurements and calculations have been performed
for cesium.
In 1989, calculations for cesium reached an accuracy of 1\%
\cite{dzuba89,blundell90}.
At this time, the experimental accuracy was at a level of 2\%.
In 1997, Wood {\it et al} measured the PNC amplitude in cesium
to an accuracy of 0.35\% \cite{wood97}.
In light of new measurements of quantities relevant to the PNC amplitude
(which are now in better agreement with the calculated values),
Bennett and Wieman analyzed the accuracy of the calculations
\cite{dzuba89,blundell90} and concluded that it is 0.4\% \cite{wieman99}.
With this accuracy, the nuclear weak charge deviates from the
Standard Model value by $2.5~\sigma$.

The suggestion that the actual accuracy of the calculations is $0.4\%$
immediately raised questions about values of small corrections which
were neglected in \cite{dzuba89,blundell90}.
It was shown by Derevianko \cite{derevianko00} that the contribution
of the Breit interaction to the PNC amplitude, $E_{PNC}$, is substantially
larger than previous estimates and reduces the deviation from the
Standard Model.
Sushkov pointed out \cite{sushkov01} that neglected radiative corrections
to $E_{PNC}$ must also be considered.

In this work we have performed a complete calculation of the $6s-7s$ PNC
amplitude in cesium.
We use the same method of calculation as that used
in the 1989 work \cite{dzuba89}
(this method is described in Section \ref{sec:method})
but with much better numerical accuracy.
The method which we call ``many-body perturbation theory in the screened
Coulomb interaction'' was developed to treat the most important
sequences of higher-order correlation diagrams in all orders.
Note that this is an all-order technique which is not a version of
the popular coupled-cluster technique.
This method has proven to be very accurate for alkaline
atoms and has produced very good results in a number of calculations.
We believe that it can hardly be improved in terms
of incorporating more correlation diagrams.
However, since more computer power is available to us now compared to
what we had in 1989 it is important to check the stability of the
results with the improved numerical accuracy.
In this work we have also calculated the Breit and radiative corrections
to the PNC amplitude. We have also accounted for the small shift in
$E_{PNC}$ due to consideration of the neutron distribution.

We list the results of our work below.

Repeating the 1989 calculation with much higher numerical accuracy,
we have obtained the same result:
$E_{PNC}=0.908$ (in units $iea_{B}(-Q_{W}/N)\times 10^{-11}$).
From the analysis below, we have concluded that the accuracy
of this value is about $0.5\%$.

Our calculation of the Breit interaction corrects $E_{PNC}$ by $-0.0055$.
This is in agreement with \cite{derevianko00,dzuba01,kozlov01}.

Radiative corrections to the weak charge
of order $\alpha$ have been calculated
in \cite{marciano} (see also \cite{lynnbed}).
However, there are important radiative corrections
of order $Z\alpha ^{2}$ and $Z^2\alpha ^{3} \ln^2(\lambda/R_n)$
which have recently been calculated in
\cite{milstein} (note that the latter correction is larger).
Here $\lambda$ is the electron Compton wavelength, and
$R$ is the nuclear radius.
Their contribution to $E_{PNC}$ is $0.004$. This contribution
originates from the radiative corrections to the weak matrix element
due to the Uehling potential (recently this contribution was calculated
in \cite{johnson01}).
We should add that there are other important contributions which have
the same magnitude.
In the sum-over-states approach (see Section \ref{sec:pnc})
they correspond to radiative corrections to the
energy intervals (-0.003) and E1 amplitudes (0.003).
Parametrically, they are proportional to
$Z^2\alpha ^{3} \ln(1/Z^2\alpha ^2)$. There are also corrections
to the weak matrix elements of order $Z^2\alpha ^{3} \ln(\lambda/R)$.
Our  estimate of these corrections shows that they have opposite
sign to the Uehling potential contribution and the same magnitude. As
a final result we suggest the following estimate for the contribution
of the radiative corrections to  $E_{PNC}$: $0.000 \pm 0.004$.

In this work we use an improved charge distribution
compared to that used in 1989.
This shifts $E_{PNC}$ very slightly, by 0.0007.

The neutron distribution in nuclei is slightly different
from the well-studied charge distribution. This produces
a small correction to $E_{PNC}$.
In this work we have found this correction to be $-0.0018$.
This is in agreement with the result of Ref. \cite{derevianko}.

This leaves us with $E_{PNC}=0.902 \pm 0.7 \%$.
Combined with measurements of $E_{PNC}/\beta$ from \cite{wood97}
and $\beta =27.00(6)a_{B}^{3}$ (the average value of the most accurate
results from \cite{dzuba00,wieman99} and \cite{dzuba97,wood97}) this gives
$Q_{W}=-72.39\Big( 1 \pm 0.4\%({\rm exp})\pm 0.7\%({\rm theory})\Big)$
which deviates by
$\Big( 1.0 \pm 0.4({\rm exp})\pm 0.7({\rm theory})\Big) \%$ from
the Standard Model value $Q_{W}= -73.09(3)$ \cite{groom00}.
From our point of view, this does not look like a significant
deviation from the Standard Model.

\section{Method of calculation}
\label{sec:method}

The method that we use here was developed in the works
\cite{dzuba89,dzuba89energy,dzuba89e1hfs}.
As was emphasized in the introduction, it is an all-order technique,
in terms of treating correlations, and is not a version of the popular
coupled-cluster method.
The dominating sequences of higher-order correlation diagrams
correspond to real physical phenomena like screening of the
Coulomb interaction and the hole-particle interaction.
They are included in all orders in our technique.
While many-body perturbation theory in the
residual Coulomb interaction does not converge, there is good convergence
of our method ``many-body perturbation theory in the
screened Coulomb interaction''.
Another strong point of the method is that its complexity does not go
beyond the calculation of energies.
The most complicated and time consuming part of the method is the
calculation of the correlation potential $\hat{\Sigma}$.
$\hat{\Sigma}$ is used to calculate single-electron Brueckner orbitals.
The calculation of matrix elements with Brueckner orbitals
is easy and they already include most of the correlations.
This makes the calculation of hyperfine structure, PNC, etc. as simple as
the calculation of energies.

The nuclear spin-independent weak interaction of an electron with
the nucleus is
\begin{equation}
\label{eq:h_w}
\hat{H}_{W}=\frac{G_{F}}{2\sqrt{2}}\rho (r)Q_{W}\gamma _{5}
\end{equation}
where $G_{F}$ is the Fermi constant, $Q_{W}$ is the weak charge of the
nucleus, $\gamma _{5}$ is a Dirac matrix, and $\rho (r)$ is the nuclear
density (note that because the nuclear weak charge in the Standard
Model is approximately equal to the number of neutrons, the
density $\rho (r)$ should be taken as the neutron density;
see Section \ref{sec:pnc}).

There are essentially two different methods by which the PNC E1 amplitude
can be calculated:
from a ``mixed-states'' approach
(in which the external fields are taken into account
in the electron orbitals) and from a ``sum-over-states'' approach
(perturbation theory sum over intermediate opposite parity states).
The most complete calculation using the mixed-states method
was performed in the work \cite{dzuba89}.
In this work all-orders summation of dominating diagrams in
the residual electron-electron interaction was included.
This is the method used in the current work.
(For the sum-over-states method, see the discussion in Section
\ref{sec:pnc}.)

The method of calculation we use is applicable for $N$-electron atoms with
one valence electron. This method is particularly effective for cesium,
compared to other heavy atoms, as in this case the external electron has
very little overlap with the tightly-bound core, enabling the use of
perturbation theory in the calculation of the residual interaction of the
external electron with the core.

The calculations start from the relativistic Hartree-Fock (RHF) method
in the $\hat{V}^{N-1}$ approximation.
The single-electron RHF Hamiltonian is
\begin{equation}
\label{eq:RHF}
\hat{H}_{0}=c\mbox{\boldmath$\alpha$}\cdot \hat{{\bf p}}+(\beta -1)c^{2}-
Z\alpha/r+\hat{V}^{N-1} \ ,
\end{equation}
$\mbox{\boldmath$\alpha$}$ and $\beta$ are Dirac matrices and
$\hat{{\bf p}}$ is the electron momentum.
For cesium, the accuracy of the RHF energies is of the order of $10\%$.

We use the time-dependent Hartree-Fock (TDHF) method
(which is equivalent to the random-phase approximation with exchange)
to calculate the interaction of external fields with atomic electrons.
In this paper we deal with two external fields:
the electric field of the photon (E1 transition amplitudes)
and the weak field of the nucleus.
In the RHF approximation the interaction between an external field
$\hat{H}_{\rm ext}$ and atomic electrons is defined by the matrix element
$\langle \psi _{2}|\hat{H}_{\rm ext}|\psi _{1}\rangle$,
where $\psi _{1}$ and $\psi _{2}$ are RHF orbitals.
Inclusion of the polarization of the atomic core by an external field is
reduced to the addition of a correction $\delta \hat{V}$
(which is the correction to the Hartree-Fock potential due to the interaction
between the core and the external field)
to the operator which describes the interaction,
$\langle \psi _{2}|\hat{H}_{\rm ext}+
\delta \hat{V}|\psi _{1}\rangle$.

The TDHF contribution to $E_{PNC}$ between states $6s$ and $7s$ in
the mixed-states approach is given by
\begin{equation}
\label{eq:TDHF}
E_{PNC}^{TDHF}=\langle \psi _{7s}|\hat{H}_{E1}+\delta \hat{V}_{E1}|
\delta \psi _{6s}\rangle +
\langle \psi _{7s}|\hat{H}_{W}+\delta \hat{V}_{W}|
X _{6s}\rangle +
\langle \psi _{7s}|\delta \hat{V}_{E1W}|\psi _{6s}\rangle  \
\end{equation}
(of course, this term can instead be expressed in terms of corrections
to $\psi _{7s}$).
Here $\delta \psi $ and $\delta \hat{V}_{W}$ denote corrections to
single-electron RHF wavefunctions and the Hartree-Fock potential caused
by the weak interaction.
These corrections are found by solving
\begin{equation}
(\hat{H}_{0}-\epsilon)\delta \psi =
-(\hat{H}_{W}+\delta \hat{V}_{W})\psi \ .
\end{equation}
The positive (negative) frequency corrections $X$ ($Y$)
due to the E1 field of the photon are found from the equations
\begin{eqnarray}
(\hat{H}_{0}-\epsilon -\omega)X &=&
-(\hat{H}_{E1}+\delta \hat{V}_{E1})\psi \\
(\hat{H}_{0}-\epsilon +\omega)Y&=&
-(\hat{H}_{E1}^{\dagger}+\delta \hat{V}_{E1}^{\dagger})\psi  \ .
\end{eqnarray}
The correction to the potential due to the E1 field is
$\delta \hat{V}_{E1}$ ($\delta \hat{V}_{E1}^{\dagger}$).
The correction $\delta \hat{V}_{E1W}$ to the core potential
is due to the simultaneous action of the
weak field and the electric field of the photon.

\begin{figure}[b]
\centerline{\psfig{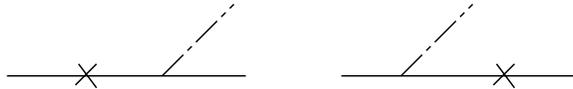}}
\caption{Lowest-order diagrams for PNC.
Solid line denotes the bound electron;
cross is the weak interaction; and dashed line is the E1 field.}
\label{fig:pnc}
\end{figure}

\begin{figure}[b]
\centerline{\psfig{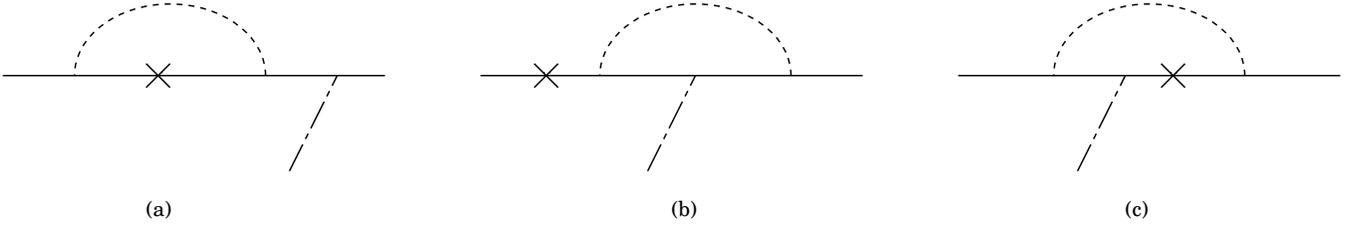}}
\caption{Examples of diagrams representing the polarization of the
atomic core by external fields.
The dashed loop is the Coulomb field.
(The diagrams we have presented are lowest-order exchange diagrams;
there are also direct diagrams.)
In diagrams (a) and (b) the core is
polarized by a single field.
Diagram (c) corresponds to the polarization of the core by both fields.}
\label{fig:corepol}
\end{figure}

The TDHF contribution to $E_{PNC}$ (Eq. \ref{eq:TDHF})
corresponds to calculating the lowest-order PNC diagrams
presented in Fig. \ref{fig:pnc} with the core polarization diagrams
(see Fig. \ref{fig:corepol}) included in all orders in the Coulomb
interaction.

\begin{figure}[b]
\centerline{\psfig{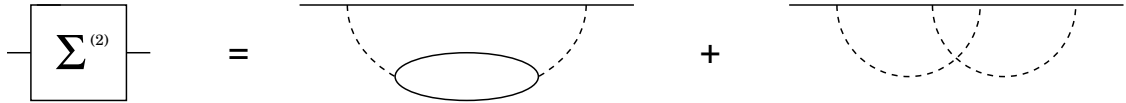}}
\caption{Second-order correlation diagrams.
Dashed line is the Coulomb interaction.
Loop is the polarization of the atomic core.}
\label{fig:2ndorder}
\end{figure}

The wavefunctions are improved by taking into account
electron-electron correlations.
The correlations are calculated using the ``correlation potential'' method
\cite{dzuba87} which corresponds to adding a non-local correlation potential
$\hat{\Sigma}$ to the potential $\hat{V}^{N-1}$ in the RHF Hamiltonian
(\ref{eq:RHF}) and then solving for the states of the external electron.
The correlation potential is defined such that its
average value coincides with the correlation correction to energy,
$\delta E_{i}=\langle\psi _{i}|\hat{\Sigma}|\psi _{i}\rangle$.
The correlation potential is calculated by means of many-body
perturbation theory in the residual Coulomb interaction
\begin{equation}
\hat{U}=\hat{H}-\sum_{i=1}^{N}\hat{H}_{0}({\bf r}_{i})=
\sum _{i<j}^{N}\frac{1}{|{\bf r}_{i}-{\bf r}_{j}|}-
\sum _{i=1}^{N}\hat{V}^{N-1}({\bf r}_{i}),
\end{equation}
where $\hat{H}$ is the exact Hamiltonian of an atom.
The lowest-order correlation diagrams $\hat{\Sigma}^{(2)}$
(second-order in $\hat{U}$) are presented in Fig.~\ref{fig:2ndorder}.

\begin{figure}[b]
\centerline{\psfig{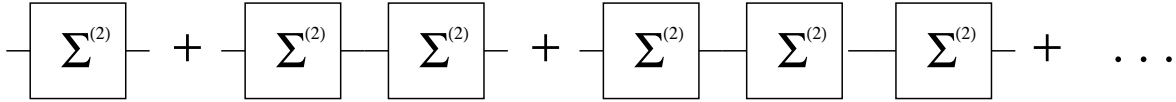}}
\caption{Chaining of the self-energy operator.}
\label{fig:sechain2}
\end{figure}

Brueckner orbitals are obtained by adding $\hat{\Sigma}^{(2)}$ to the
Hartree-Fock potential $\hat{V}^{N-1}$ and solving the equation
\begin{equation}
(\hat{H}_{0}+\hat{\Sigma}^{(2)}-\epsilon)\psi =0
\end{equation}
iteratively for the states of the external electron.
This corresponds to chaining the self-energy operator to all orders
(Fig.~\ref{fig:sechain2}).
This chain of diagrams is enhanced by the small denominator,
corresponding to the excitation energy of an external electron
(in comparison to the excitation of a core electron).
At this level of calculation (with ``bare'' $\hat{\Sigma}$)
the accuracy for energy levels for Cs is about 1\%.

\begin{figure}[b]
\centerline{\psfig{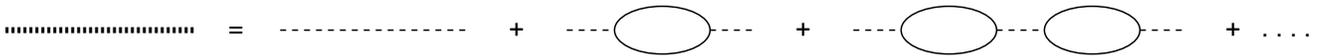}}
\caption{Screening of the Coulomb interaction.}
\label{fig:screening}
\end{figure}

\begin{figure}[b]
\centerline{\psfig{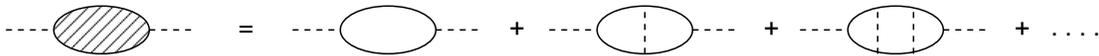}}
\caption{Hole-particle interaction in the polarization operator.}
\label{fig:hpchain}
\end{figure}

Using the correlation potential method and the Feynman diagram technique
we include into $\hat{\Sigma} ^{(2)}$ two series of dominating higher
order diagrams which are calculated in all orders of perturbation theory
\cite{dzuba89,dzuba89energy,dzuba89e1hfs}.
These are screening of the electron-electron interaction
(Fig.~\ref{fig:screening}) and the hole-particle interaction
(Fig.~\ref{fig:hpchain}).
The electron-electron screening is a collective phenomenon.
The corresponding chain of diagrams is enhanced by a factor
approximately equal to the number of electrons in the external
closed subshell (the $5p$ electrons in Cs).
The hole-particle interaction is enhanced by the large
zero multipolarity diagonal matrix elements of the
Coulomb interaction.
This interaction accounts for the discrete spectra in noble gases.
We will denote the dressed self-energy operator
by $\hat{\Sigma}$ (Fig.~\ref{fig:hpscreense}).
With this $\hat{\Sigma}$ the Brueckner energies for Cs have an accuracy
of the order of $0.1\%$.

The wavefunctions can be further modified by placing a coefficient before
$\hat{\Sigma}$ such that the corresponding energy coincides with the
experimental value. This fitting of the Brueckner orbitals can be considered
as a way of including other higher-order diagrams into the calculations.

\begin{figure}[b]
\centerline{\psfig{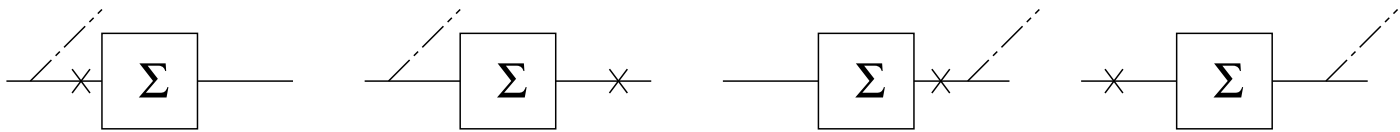}}
\caption{Lowest-order correlation corrections to the PNC E1 transition
amplitude.}
\label{fig:pncdom}
\end{figure}

If we use Brueckner orbitals instead of RHF orbitals to calculate
the PNC amplitude in Eq. \ref{eq:TDHF}, then we include all-orders
in $\hat{\Sigma}$ contributions to $E_{PNC}$.
However, the correlation potential is energy-dependent,
$\hat{\Sigma}=\hat{\Sigma}(\epsilon)$.
So, in first order, we should consider the proper energy dependence.
The first-order in $\hat{\Sigma}$ corrections to $E_{PNC}$ are
presented diagrammatically in Fig. \ref{fig:pncdom}.
We can write these as
\begin{equation}
\langle \psi _{7s}|\hat{\Sigma}(\epsilon _{7s})|\delta X_{6s}\rangle
+\langle \delta\psi _{7s}|\hat{\Sigma}(\epsilon _{7s})|X_{6s}\rangle
+\langle \delta Y_{7s}|\hat{\Sigma}(\epsilon _{6s})|\psi_{6s}\rangle
+\langle Y_{7s}|\hat{\Sigma}(\epsilon _{6s})|\delta \psi_{6s}\rangle \ .
\end{equation}
The non-linear in $\hat{\Sigma}$ contribution can be found by subtracting
from the all-orders result the first-order value found in the same method.

\begin{figure}[b]
\centerline{\psfig{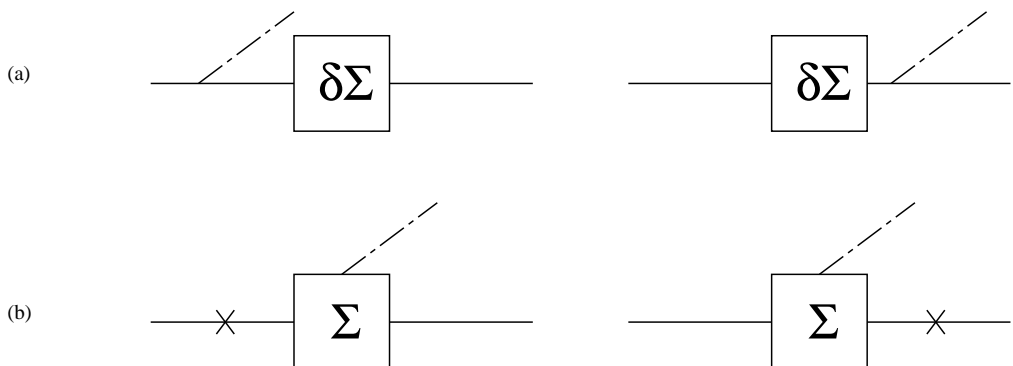}}
\caption{Small corrections to the PNC E1 transition amplitude:
external field inside the correlation potential.
In diagrams (a) the weak interaction is inside the
correlation potential ($\delta \hat{\Sigma}$ denotes the change in
$\hat{\Sigma}$ due to the weak interaction);
this is known as the weak correlation potential.
Diagrams (b) represent structural radiation
(photon field inside the correlation potential).}
\label{fig:pncint}
\end{figure}

The correlation corrections to $E_{PNC}$ we have considered so far
are usually called ``Brueckner-type'' corrections.
(In this case the external field interacts with the external electron
lines.) There are also contributions to $E_{PNC}$ in which the
external field acts inside the correlation potential
(see Fig. \ref{fig:pncint}).
Those diagrams in which the E1 interaction occurs in the internal
lines are known as ``structural radiation'',
while those in which the weak interaction occurs in the internal lines are
known as the ``weak correlation potential''.
There is another second-order correction to the amplitudes which arises
from the normalization of states \cite{dzuba87}.
The structural radiation, weak correlation potential, and
normalization contributions are suppressed by the small parameter
$E_{\rm ext}/E_{\rm core}\sim 1/10$, where  $E_{\rm ext}$ and
$E_{\rm int}$ are excitation energies of the external and core electrons,
respectively.

In this work we also calculate the contribution of the Breit
interaction.
We do not discuss the method of calculation here,
but refer the interested reader to the works \cite{dzuba01,dzuba}.

We postpone the discussion of our method for the calculation of radiative
corrections to Section \ref{sec:rad}.

\section{$6s-7s$ PNC amplitude}
\label{sec:pnc}

The results of our calculation for the  $6s-7s$ PNC amplitude
are presented in Table~\ref{tab:pnci}.
Notice that the time-dependent Hartree-Fock value gives a
contribution to the total amplitude of about $98\%$.
The point is that there is a strong cancellation of the
correlation corrections to the PNC amplitude.
The stability of the PNC amplitude compared to other
quantities in which the correlation corrections are large
will be discussed in more detail in Section \ref{sec:accuracy}.
Notice that the values are in agreement with our 1989 result
$0.908\times 10^{-11}iea_{B}(-Q_{W}/N)$
(see ``Subtotal'' of Table \ref{tab:pnci} for the current calculation).
The higher numerical accuracy of the current work has therefore
not changed the previous result.

The mixed-states approach has also been performed in
\cite{blundell90} and \cite{kozlov01} to determine the PNC amplitude
in cesium.
However, in these works the screening of the
electron-electron interaction was included in a simplified way.
In \cite{blundell90} empirical screening factors were placed before
the second-order correlation corrections $\hat{\Sigma}^{(2)}$ to fit the
experimental values of energies.
Kozlov and Porsev introduced screening factors based on average screening
factors calculated for the Coulomb integrals between valence electron
states.
The results obtained by these groups
(without the Breit interaction, i.e., corresponding to the
Subtotal of Table \ref{tab:pnci}),
$0.904$ \cite{blundell90} and $0.905$ \cite{kozlov01},
are close to ours.
To be sure that we understand this difference, we performed a
pure second-order (i.e., using $\hat{\Sigma}^{(2)}$)
calculation and fitted the energies (as was done
in \cite{blundell90}) and reproduced their result, $0.905$.

In the work \cite{blundell90} a calculation using the
sum-over-states method was also performed.
In the sum-over-states approach the $6s-7s$ PNC amplitude
is expressed in the form
\begin{equation} \label{sum}
E_{PNC}=\sum _{n} \Big(
\frac{\langle 7s |\hat{H}_{W}|np\rangle \langle np|\hat{H}_{E1}|6s \rangle}
{E_{7s}-E_{np}} +
\frac{\langle 7s |\hat{H}_{E1}|np\rangle \langle np|\hat{H}_{W}|6s \rangle}
{E_{6s}-E_{np}} \Big) \ .
\end{equation}
The authors of reference \cite{blundell90} include single, double, and
selected triple excitations into their wavefunctions.
Note, however, that even if wavefunctions of $6s$, $7s$, and intermediate
$np$ states are calculated exactly
(i.e., with all configuration mixing included)
there are still some missed contributions in this approach.
Consider, e.g., the intermediate state $6p\equiv 5p^{6}6p$.
It contains an admixture of states $5p^{5}ns6d$:
$\tilde{6p}=5p^{6}6p + \alpha 5p^{5}ns6d+...$.
This mixed state is included into the  sum (\ref{sum}).
However, the sum (\ref{sum}) must include all many-body states
of opposite parity.
This means that the state $\tilde{5p^{5}ns6d}=
5p^{5}ns6d- \alpha 5p^{6}6p+...$ should also be included into
the sum. Such contributions to $E_{PNC}$ have never
been estimated directly within the sum-over-states approach.
However, they are included into the mixed-states calculation.
The result of the sum-over-states approach, 0.909,
is very close to the result of the mixed-states aproach, 0.908.
It is important to note that the omitted higher-order many-body corrections
are different in these two methods.
This may be considered as an argument that the omitted many-body corrections
in both calculations are small.
Of course, here we assume that the omitted many-body corrections to both
values (which, in principle, are completely different) do not
``conspire'' to give exactly the same magnitude.

Therefore we will take $0.908$ for the value of $E_{PNC}$
(Subtotal of Table \ref{tab:pnci}) as this corresponds
to the most complete mixed-states calculation
and is in agreement with the sum-over-states calculation of
reference \cite{blundell90}.

With Breit, our result becomes $0.903\times 10^{-11}iea_{B}(-Q_{W}/N)$.
This correction is in agreement with
\cite{derevianko00,dzuba01,kozlov01}.

We use the two-parameter Fermi model for the proton and neutron distributions:
\begin{equation}
\rho (r)=\rho _{0}\Big[ 1+\exp [(r-c)/a] \Big] ^{-1}\ ,
\end{equation}
where $t=a(4\ln 3)$ is the skin-thickness,
$c$ is the half-density radius, and
$\rho _{0}$ is found from the normalization condition $\int \rho dV=1$.
In 1989 the thickness and half-density radius for the proton
distribution were taken to be $t_{p}=2.5~{\rm fm}$ and
$c_{p}=5.6149~{\rm fm}$ (corresponding to a root-mean-square
(rms) radius $\langle r_{p}^{2}\rangle ^{1/2}=4.836~{\rm fm}$).
In this work we have used improved parameters
$t_{p}=2.3~{\rm fm}$ and $c_{p}=5.6710~{\rm fm}$
($\langle r_{p}^{2}\rangle ^{1/2}=4.804~{\rm fm}$) \cite{fricke95}.
This changes the wavefunctions slightly,
leading to a very small correction to the PNC amplitude
of $0.08\%$ ($0.0007$).
(This is in agreement with a simple analytical estimate:
the factor accounting for the change in the electron density
is $\sim (4.804/4.836)^{-Z^{2}\alpha ^{2}}\sim 0.1\% \ $.)
In the work \cite{dzuba89} we used the proton distribution in the
weak interaction Hamiltonian (Eq. \ref{eq:h_w}).
In the current work we have found the small correction to
$E_{PNC}$ which arises from taking the (poorly understood)
neutron density in Eq. \ref{eq:h_w}.
We use the result of Ref. \cite{r_{np}} for the difference
$\Delta r_{np}=0.13(4)~{\rm fm}$
in the root-mean-square radii of the neutrons
$\langle r_{n}^{2}\rangle ^{1/2}$ and protons
$\langle r_{p}^{2}\rangle ^{1/2}$.
We have considered three cases which correspond to the same value of
$\langle r_{n}^{2}\rangle$: (i) $c_{n}=c_{p}$, $a_{n}>a_{p}$;
(ii) $c_{n}>c_{p}$, $a_{n}>a_{p}$; and (iii) $c_{n}>c_{p}$, $a_{n}=a_{p}$
(using the relation $\langle r_{n}^{2}\rangle \approx \frac{3}{5}c_{n}^{2}
+\frac{7}{5}\pi ^{2}a_{n}^{2}$).
We have found that $E_{PNC}$ shifts from $-0.18\%$ to $-0.21\%$
when moving from the extreme $c_{n}=c_{p}$ to the extreme $a_{n}=a_{p}$.
Therefore, $E_{PNC}$ changes by about $-0.2\%$ ($-0.0018$)
due to consideration of the neutron distribution.
This is in agreement with Derevianko's estimate,
$-0.19(8)\%$ \cite{derevianko}.

In the next section we discuss the radiative corrections to
$E_{PNC}$. Our estimate of these corrections does not shift
the PNC amplitude.

Therefore, we have
\begin{equation}
E_{PNC}=0.902 \times 10^{-11}iea_{B}(-Q_{W}/N)
\end{equation}
as our central point for the PNC amplitude.
The error will be estimated in the following sections.

\section{QED-type radiative corrections to energy levels,
wavefunctions, and the PNC amplitude}
\label{sec:rad}

The radiative corrections to the weak charge $Q_W$ have been
calculated for the free electron. However, an electron in a heavy
atom is bound, and this produces additional radiative
corrections proportional to $\alpha (Z \alpha)^n$,
$n=1,2,...$. Recently such corrections were considered
by Milstein and Sushkov \cite{milstein}. They found that
the most important are corrections enhanced by
the large parameter $\ln(\lambda/R)$, where $\lambda=\hbar/mc$
is the electron Compton wavelength and $R$ is the nuclear
radius. This type of correction arises from the radiative
corrections to the electron wavefunction near the nucleus.
In this region the $s$-wave and $p_{1/2}$-wave (lower Dirac component)
electron densities are singular, $|\psi(r)|^2 \sim r^{-Z^2\alpha^2}$.
The radiative corrections modify the potential at small distances
$r<\lambda$, ${\tilde V }(r)= -Z \alpha (1+\delta)/r$.
Correspondingly, the electron wavefunctions change,
$|\psi(r)|^2 \sim r^{-Z^2\alpha^2 (1 +\delta)^2}$ for $r<\lambda$.
This gives the radiative correction factor for the electron
density inside the nucleus,
\begin{equation}
\label{psirad}
\frac{|\psi(R)|^2}{|\psi(\lambda)|^2} \sim \Big( \frac{\lambda}{R}\Big)
^{Z^2\alpha^2 2\delta}=\exp \Big( 2\delta Z^2\alpha^2 \ln(\lambda /R) \Big)
\ .
\end{equation}
For the Uehling (vacuum polarization) potential
$\delta \sim \alpha \ln(\lambda /r)$ \cite{berestetskii}.
This gives an additional power of the large  parameter $\ln(\lambda / R)$.
This leads Milstein and Sushkov \cite{milstein} to conclude that the
Uehling potential gives a dominating radiative correction
to $E_{PNC}$, $\sim  Z^2\alpha^3 \ln^2(\lambda/ R)$.
Numerical calculations of the Uehling potential contribution
have been performed in \cite{johnson01} and in the present work.
This radiative correction increases $E_{PNC}$ by 0.4\%.

Milstein and Sushkov \cite{milstein} demonstrated that there are no other
radiative corrections which are enhanced by  $\ln^2(\lambda /R)$.
However, any correction to the potential with nonzero
$\delta(R) \sim \alpha$ gives a correction to the electron density
$\sim  Z^2\alpha^3 \ln (\lambda /R)$. We demonstrate below
that such corrections are also important and give a contribution
of opposite sign to that of the Uehling potential.

Let us start our discussion from the radiative corrections to energy
levels (the Lamb shift).
The calculation of the shift can be divided into two parts:
one in which the electron interaction with virtual photons of
high-frequency are considered, and one in which
virtual photons of low-frequency are considered.

\begin{figure}[b]
\centerline{\psfig{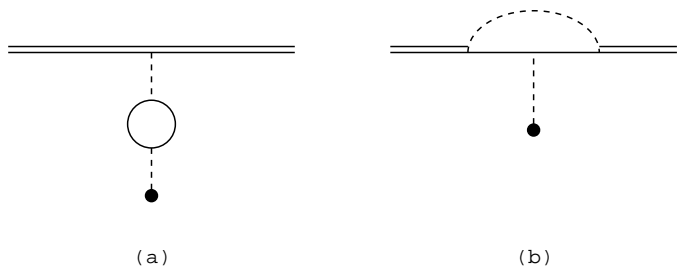}}
\caption{High-frequency contribution to radiative corrections.
Diagram (a) corresponds to the Uehling potential.
Diagram (b) is the vertex correction.
The double line represents the bound electron;
the single solid line is the free electron;
the Coulomb interaction is denoted by the dashed line;
and the filled circle denotes the nucleus.}
\label{fig:rad}
\end{figure}

In the high-frequency case the external field
(the strong nuclear Coulomb field) need only be included
to first order.
In this case the contributions to the Lamb shift arise
from the diagrams presented in Fig. \ref{fig:rad}.
The contribution of the Uehling potential (Fig. \ref{fig:rad}(a))
to the Lamb shift is very small.
The main contribution comes from the vertex correction
(Fig. \ref{fig:rad}(b)).
(In the case of a free electron the vertex diagrams give the electric
$f(q^{2})$ and magnetic $g(q^{2})$ formfactors.)
The perturbation theory expression for $f(q^2)$ contains an
infra-red divergence and requires a low-frequency
cut-off parameter $\kappa$ - see, e.g., \cite{berestetskii}.
Assuming $q^2 \ll m^2 c^2$, the high-frequency contribution to the
Lamb shift can be presented as a potential given by the following
expression \cite{berestetskii}
\begin{eqnarray}
\delta \Phi ({\bf r})
&=&\Big[ \delta \Phi _{f}+\delta \Phi _{U}\Big] +\delta \Phi _{g}\nonumber \\
&=&\frac{\alpha \hbar ^{2}}{3\pi m^{2}c^{2}}\Big(
\ln \frac{m}{2\kappa} +\frac{11}{24}-\frac{1}{5}\Big)
\Delta \Phi ({\bf r})-
i\frac{\alpha \hbar}{4\pi mc}\mbox{\boldmath$\gamma$}\cdot
\mbox{\boldmath$\nabla$}\Phi ({\bf r}) \ .
\end{eqnarray}
For the Coulomb potential, $\Delta \Phi =-4\pi Ze\delta ({\bf r})$.
Here the last long-range term ($\delta \Phi _{g}$) comes from
the anomalous electron magnetic moment ($g(0)$);
the infra-red cut-off parameter $\kappa$ appears from the electric
formfactor $f(q^{2})$.
This term with the large $\ln \frac{m}{2\kappa}$ gives the dominant
contribution to the Lamb shift of $s$-levels.
The infra-red divergence for $\kappa \rightarrow 0$ indicates the
importance of the low-frequency contribution for this term.

If we go beyond the approximation
$q^{2}<<m^{2}c^{2}$, the term $\delta \Phi _{f}$ should be
associated with a non-local self-energy operator
${\hat \Sigma}_{\rm rad} ({\bf r},{\bf r}',E)$ with typical values
$|{\bf r}-{\bf r}'|\lesssim \frac{\hbar}{mc}$
and $r\sim r' \lesssim \frac{\hbar}{mc}$.
We need this operator in a simple limit, $E<<mc^{2}$.
In this case we can approximate it by a two-parametric $(A,b)$ potential
\begin{equation}
\delta \Phi _{f}=-A\frac{\alpha}{\pi} {\rm e}^{-b\frac{mc}{\hbar}r}\Phi \ .
\end{equation}
The parameters $A$ and $b$ in $\delta \Phi _{f}$ can be found from the
fit of the Lamb-shift of the high Coulomb levels $3s$, $4s$, $5s$ and
$3p$, $4p$ and $5p$ (in one-electron ions) which were calculated as
a function of the nuclear charge $Z$ in Refs. \cite{mohr}.
We have checked that $A=1.25$ and $b=1$ fit all these
Lamb shifts quite accurately (we have found slightly different potential
strengths for $s$- and $p$-waves, $A_s=1.17$ and $A_p=1.33$).
The anomalous magnetic moment contribution is
\begin{equation}
\delta \Phi _{g}=-i\frac{\alpha \hbar}{4\pi mc}\mbox{\boldmath$\gamma$}
\cdot \mbox{\boldmath$\nabla$}\Phi \ .
\end{equation}
The Uehling potential for a finite nucleus is given by \cite{fullerton76}
(in atomic units ($\hbar =m=e=1$, $\alpha =1/c$))
\begin{equation}
\delta \Phi _{U}=-\frac{2\alpha ^{2}}{3r}\int _{0}^{\infty}dx~
x\rho(x)\int _{1}^{\infty}dt~\sqrt{t^{2}-1}
\Big(
\frac{1}{t^{3}}+\frac{1}{2t^{5}}\Big)
\Big(
{\rm e}^{-2t|r-x|/\alpha} -{\rm e}^{-2t(r+x)/\alpha} \Big)    \ ,
\end{equation}
where $\rho (x)$ is the nuclear charge density.
It is more convenient to use a simpler formula for
$\delta \Phi _{U}$ for $r\geq R$, $R$ is the nuclear radius,
\begin{eqnarray}
&&\delta \Phi _{U}(r)= \Phi (r)\frac{\alpha ^{4}}{8\pi R^{3}}
\int _{1}^{\infty}dt~ \sqrt{t^{2}-1}
\Big(
\frac{1}{t^{5}}+\frac{1}{2t^{7}}\Big)
{\rm e}^\frac{-2tr}{\alpha}I(x)  \ , \\
&&I(x)=-{\rm e}^{x}+{\rm e}^{-x}+x{\rm e}^{x} +x{\rm e}^{-x} \ ,
\qquad x=2tR/\alpha \ ,
\end{eqnarray}
and take $\delta \Phi _{U}(r<R)=\delta \Phi _{U}(r=R)$.
There is practically no loss of numerical accuracy in this
approximation since a typical scale for the variation of
$\delta \Phi _{U}(r)$ is given by the electron compton length
$\frac{\hbar}{mc}>>R$.
The radiative corrections for Cs energy levels are presented
in Table \ref{tab:radenergies}.

Note that for the most important term $\delta \Phi _{f}$ we do not
use the assumption $Z\alpha <<1$ since we fitted the exact results
for the single-electron ions.
The contribution of the Uehling potential $\delta \Phi _{U}$ to the Lamb
shift is always small.
Also, Milstein and Strakhovenko have shown in Ref. \cite{milstein83}
that higher $Z\alpha$ corrections are numerically not important
for this potential.
The potential $\delta \Phi _{g}$ due to the magnetic formfactor
is a long-range one.
This also hints that there are no large higher $Z\alpha$
corrections here.

We can use $\delta \Phi$ to estimate the contribution of
QED-type radiative corrections to the electron wavefunction
and  PNC amplitude $E_{PNC}$.
Note that it is not enough to calculate the radiative corrections to
the matrix element of the weak interaction
$\langle n'p_{1/2}|\hat{H}_{W}|ns\rangle$.
Corrections to the energy intervals like $6s-6p$  are also important
since these intervals are small at the scale of the atomic unit
($ \sim 1/20$) and sensitive to perturbations.
The change in the energies also influences the large-distance behavior
of the electron wavefunctions which determine the usual
E1 amplitudes in the sum-over-states approach.
Thus, the simplest way to proceed is to include $\delta \Phi$ into the
Dirac-Hartree-Fock equations and then perform all calculations.

The results are the following:
the Uehling potential $\delta \Phi _{U}$ increases $E_{PNC}$ by
$0.41\%$ (in agreement with \cite{johnson01});
the contribution of $\delta \Phi _{g}$ is small, $-0.03 \%$
(due to cancellation of the contributions of the corrections
to the $s$-wave and $p$-wave);
and the contribution of $\delta \Phi _{f}$ is $-0.65\%$.
The total result of $\delta \Phi$ is $-0.27\%$
(and opposite to that of the Uehling potential).

Note that the replacement of
${\hat \Sigma}_{\rm rad} ({\bf r},{\bf r}',E=0)$ by a
local parametric potential ($\delta \Phi _{f}$) may be quite a crude
approximation. Therefore, we have tested the sensitivity of
the result to variation of the parameter $b$
(parameter $A$ is a function of $b$, it is found from the fit
of the Coulomb energy levels).
The increase of the radius of the potential $\delta \Phi _{f}$ two times
($b=0.5$) increases $E_{PNC}$ by $0.3\%$
(the total contribution of $\delta \Phi$ is $0.04\%$).

We do not discuss here the $Z\alpha$ correction to the
$Z$-boson exchange vertex. Milstein and Sushkov \cite{milstein} have
shown that this correction does not contain the large
parameter $\ln(\lambda /R)$ if the calculations
are performed in the Landau gauge and they concluded
that this correction should be small.

Here we should note that the radiative corrections to
the wavefunctions (as well as the wavefunctions themselves)
are not gauge invariant.
However, the change of the electron density inside the nucleus is
a gauge invariant, observable phenomenon.
This makes the consideration presented above meaningful.

Of course, our estimate of QED-type radiative corrections to
$E_{PNC}$ should not be considered as an accurate calculation.
The aim is to show that the consideration of just the single
Uehling potential contribution to the radiative corrections
can give the wrong impression.
Indeed, this contribution to $E_{PNC}$ is enhanced by the large
parameter $\ln^2(\lambda /R)$.
However, this only makes it comparable to other contributions which
in the case of energy levels were an order of magnitude
larger than that of the Uehling potential.
A conservative estimate of the radiative correction contribution
can be presented as $0.0 \pm 0.4 \%$.
This range covers our two values (-0.27\% and 0.04\%) as well as
the value 0.4\% obtained in \cite{milstein,johnson01}.


\section{Estimate of accuracy of PNC amplitude}
\label{sec:accuracy}

We have estimated the error of the PNC amplitude in a number
of different ways. There are two main methods:
(i) root-mean-square (rms) deviation of calculated energy intervals,
E1 amplitudes, and hyperfine structure (hfs) constants
from the accurate experimental values;
(ii) influence of fitting of energies and hyperfine structure
constants on the PNC amplitude.

\subsection{Root-mean-square deviation}

Remember that the PNC amplitude can be expressed as a sum over
intermediate states (see formula \ref{sum}).
Each term in the sum is a product of E1 transition amplitudes,
weak matrix elements, and energy denominators.
There are three dominating contributions to the $6s-7s$ PNC amplitude
in Cs \cite{blundell90}:
\begin{eqnarray}
E_{PNC}&=&
\label{sumcs}
\frac{\langle 7s|\hat{H}_{E1}|6p\rangle \langle 6p|\hat{H}_{W}|6s\rangle}
{E_{6s}-E_{6p}} +
\frac{\langle 7s|\hat{H}_{W}|6p\rangle \langle 6p|\hat{H}_{E1}|6s\rangle}
{E_{7s}-E_{6p}}+
\frac{\langle 7s|\hat{H}_{E1}|7p\rangle \langle 7p|\hat{H}_{W}|6s\rangle}
{E_{6s}-E_{7p}}+...
\nonumber \\
&=& -1.908+1.493+1.352+...=0.937 +... \ .
\end{eqnarray}
While we do not use the sum-over-states approach in our calculation of
the PNC amplitude, it is instructive to analyze the accuracy of
the E1 transition amplitudes, weak matrix elements, and
energy intervals which contribute to Eq. \ref{sumcs} as
they have been calculated using the same method as that used to
calculate $E_{PNC}$.

Let us begin with the energy intervals.
The calculated ionization energies are presented in Table \ref{tab:energies}.
The Hartree-Fock values deviate from experiment by $10\%$.
Including the second-order correlation corrections ${\hat \Sigma} ^{(2)}$
reduces the error to $\sim 1\%$.
When screening and the hole-particle interaction are included into
${\hat \Sigma} ^{(2)}$ in all orders,
the energies improve, $\sim 0.1\%$.
The percentage deviations from experiment of the energy intervals of interest
are: $E_{6s}-E_{6p}$, $-0.3$; $E_{7s}-E_{6p}$, $0.8$;
and $E_{6s}-E_{7p}$, $-0.01$. The rms error is $0.5\%$.
We can in fact reproduce energy intervals exactly by placing coefficients
before the correlation potential. Because this fitting of the
energies appears to improve the wavefunctions (e.g., electromagnetic
amplitudes and hyperfine structure constants improve)
we will use this procedure in the following estimates.

The relevant radial integrals (E1 transition amplitudes) are presented in
Table \ref{tab:e1i}.
These were calculated with the energy-fitted ``bare'' correlation
potential $\hat{\Sigma}^{(2)}$ and the (unfitted and fitted)
``dressed'' potential $\hat{\Sigma}$.
Structural radiation and normalization contributions were also
included.
In Table \ref{tab:e1ii} the percentage deviations of
the calculated values from experiment are listed.
Without energy fitting, the rms error is $0.3\%$.
Fitting the energy improves the accuracy, $\sim 0.1\%$.

We cannot directly compare weak matrix elements with experiment.
Like the weak matrix elements, hyperfine structure is determined
by the electron wavefunctions in the vicinity of the nucleus,
and this is known very accurately.
The hyperfine structure constants calculated in different approximations
are presented in Table \ref{tab:hfsi}.
Corrections due to the Breit interaction, structural radiation,  and
normalization are included.
The percentage deviations from experiment are shown in
Table \ref{tab:hfsii}.
The rms deviation of the calculated hfs values from experiment
using unfitted ${\hat \Sigma}$ is $1.1\%$.
With fitting, the rms error in the pure second-order approximation is $0.3\%$;
with higher orders we get $0.6\%$.
We are, however, trying to estimate the accuracy of the $s-p$ weak
matrix elements. It makes more sense for us to use the square-root
formula, $\sqrt{{\rm hfs}(s){\rm hfs}(p)}$. The errors are presented in Table
\ref{tab:hfsii}.
Notice that by using this approach the error is much smaller.
Without energy fitting, the rms error is $0.4\%$.
With fitting, the rms error in the second-order calculation
($\hat{\Sigma} ^{(2)}$) and full calculation
($\hat{\Sigma}$) is $0.2\%$.

From this section we can conclude that the rms error for the relevant
parameters is somewhere in the range $0.2-0.5\%$.

Note that from this analysis the error for the sum-over-states
calculation of $E_{PNC}$ would be larger than this, as
the errors for the energies, hfs constants, and E1 amplitudes
contribute to each of the three terms in Eq. \ref{sumcs}.
However, in the mixed-states approach, the errors do not add in this
way.
We get a better indication of the error of our calculation
of $E_{PNC}$ in the next section.

\subsection{Influence of fitting on the PNC amplitude}

In the section above we presented calculations in three different
approximations:
with unfitted $\hat{\Sigma}$,
and with $\hat{\Sigma}^{(2)}$ and $\hat{\Sigma}$ fitted with
coefficients to reproduce experimental ionization energies.
The errors in these approximations are of different magnitudes and signs.
We now calculate the PNC amplitude using these three approximations.
The spread of the results can be used to estimate the error.

The results are listed in Table \ref{tab:pncii}.
It can be seen that the PNC amplitude is very stable.

The PNC amplitude is much more stable than hyperfine structure.
This can be explained by the much smaller correlation corrections
to PNC (compare Table \ref{tab:pnci} with Table \ref{tab:hfsi}).
The stability of $E_{PNC}$ may be compared to the stability of the
usual electromagnetic amplitudes where the error is very small
(even without fitting).

We have also considered the fitting of hyperfine structure using
different coefficients before each $\hat{\Sigma}$. Using the resulting
wavefunctions, the PNC amplitude increased by $0.5\%$.
This is the maximum deviation we have obtained. We will therefore
use this as the estimate for the accuracy of the $E_{PNC}$ calculation.


\section{Conclusion}

We have obtained the result
\begin{equation}
E_{PNC}=0.902(6) \times 10^{-11}iea_{B}(-Q_{W}/N)
\end{equation}
for our calculation of the $6s-7s$ PNC amplitude in Cs.
This is in agreement with other PNC calculations,
however we would like to emphasize that our calculation is
the most complete.
The most precise measurement of the $6s-7s$ PNC amplitude in Cs
is \cite{wood97}
\begin{equation}
-\frac{{\rm Im} (E_{PNC})}{\beta}=1.5939(56)\frac{\rm mV}{\rm cm} \ ,
\end{equation}
where $\beta$ is the vector transition polarizability.
For $\beta$ we use the value
\begin{equation}
\beta =27.00(6) a_{B}^{3}
\end{equation}
which is the average value of the most accurate
results \cite{dzuba00,wieman99} and \cite{dzuba97,wood97}.
Using the conversion $|e|/a_{B}^{2}=5.1422\times 10^{12}{\rm mV}/{\rm cm}$,
we therefore obtain for the weak charge of the Cs nucleus:
\begin{equation}
Q_{W}=-72.39(29)_{\rm exp}(51)_{\rm theory} \ ,
\end{equation}
where the experimental error is obtained by adding in quadrature the
error for $\beta$ and the error for ${\rm Im}(E_{PNC})/\beta$.
This result deviates by
$\Big( 1.0 \pm 0.4({\rm exp})\pm 0.7({\rm theory})\Big) \%$ from
the Standard Model value $Q_{W}= -73.09(3)$ \cite{groom00}.
This does not represent a significant deviation.

\acknowledgments

We are grateful to A. Milstein, O. Sushkov, and M. Kuchiev for
useful discussions.
This work was supported by the Australian Research Council.


\begin{table}
\caption{Contributions to the $6s-7s$ $E_{PNC}$ amplitude
for Cs in units $10^{-11}iea_{B}(-Q_{W}/N)$.
($\hat{\Sigma}$ corresponds to the (unfitted) ``dressed''
self-energy operator.) }
\label{tab:pnci}
\begin{tabular}{ld}
TDHF & 0.8885 \\
$\langle \psi _{7s}|\hat{\Sigma}|\delta X_{6s}\rangle$ & 0.0705 \\
$\langle \delta\psi _{7s}|\hat{\Sigma}|X_{6s}\rangle$ & 0.1857 \\
$\langle \delta Y_{7s}|\hat{\Sigma}|\psi_{6s}\rangle$ & -0.0760 \\
$\langle Y_{7s}|\hat{\Sigma}|\delta \psi_{6s}\rangle$ & -0.1420 \\
Nonlinear in $\hat{\Sigma}$ correction & -0.0198 \\
Weak correlation potential & 0.0038 \\
Structural radiation & 0.0025 \\
Normalization & -0.0049 \\
 & \\
Subtotal & 0.9084 \\
 & \\
Breit & -0.0055 \\
New proton distribution & 0.0007 \\
Neutron distribution & -0.0018 \\
Radiative corrections & 0.0\\
 & \\
Total & 0.902 \\
\end{tabular}
\end{table}
\begin{table}
\caption{Radiative corrections to RHF ionization energies; units $-$cm$^{-1}$.
(See also Table \ref{tab:energies}.) }
\label{tab:radenergies}
\begin{tabular}{dddd}
$6s$ & $7s$ & $6p_{1/2}$ & $7p_{1/2}$ \\
\hline

-18.4 & -5.0 & 0.88 &
0.31 \\
\end{tabular}
\end{table}
\begin{table}
\caption{Ionization energies for Cs in units $-$cm$^{-1}$.}
\label{tab:energies}
\begin{tabular}{lllll}
State & RHF & $\hat{\Sigma} ^{(2)}$ & $\hat{\Sigma}$ &
Experiment \tablenotemark[1]\\
\hline
$6s$ & 27954 & 32415 & 31420 & 31407 \\
$7s$ & 12112 & 13070 & 12863 & 12871 \\
$6p_{1/2}$ & 18790 & 20539 & 20276 & 20228 \\
$7p_{1/2}$ & 9223 & 9731 & 9657 & 9641 \\
\end{tabular}
\tablenotetext[1]{Taken from \cite{moore}.}
\end{table}
\begin{table}
\caption{Radial integrals of E1 transition amplitudes for Cs in
different approximations. The experimental values are listed in
the last column. (a.u.)}
\label{tab:e1i}
\begin{tabular}{ldddddd}
Transition & RHF& TDHF & $\hat{\Sigma} ^{(2)}$ &
$\hat{\Sigma}$ & $\hat{\Sigma}$ & Experiment \\
       &      &              & with fitting &           & with fitting &
 \\
\hline
$6s-6p$ & 6.464 & 6.093 & 5.499 & 5.510 & 5.512 &
5.497(8) \tablenotemark[1] \\
$7s-6p$ & 5.405 & 5.450 & 5.198 & 5.165 & 5.201 &
5.185(27) \tablenotemark[2]\\
$7s-7p$ & 13.483 & 13.376 & 12.602 & 12.641 & 12.612 &
12.625(18) \tablenotemark[3] \\
\end{tabular}
\tablenotetext[1]{Ref. \cite{rafac99}.}
\tablenotetext[2]{Ref. \cite{bouchiat84}.}
\tablenotetext[3]{Ref. \cite{bennett99}.}
\end{table}
\begin{table}
\caption{Percentage deviation from experiment of calculated radial integrals
in different approximations.}
\label{tab:e1ii}
\begin{tabular}{lddd}
Transition &
\multicolumn{3}{c}{Percentage deviation} \\
 & $\hat{\Sigma}^{(2)}$ & $\hat{\Sigma}$ & $\hat{\Sigma}$ \\
 & with fitting    &          & with fitting \\
\hline
$6s-6p$ & 0.04 & 0.2 & 0.3 \\
$7s-6p$ & 0.3 & -0.4 & 0.3 \\
$7s-7p$ & -0.2 & 0.1 & -0.1 \\
\end{tabular}
\end{table}
\begin{table}
\caption{Calculations of the hyperfine structure of Cs
in different approximations. In the last column the experimental
values are listed. Units: MHz.}
\label{tab:hfsi}
\begin{tabular}{ldddddd}
State & RHF & TDHF & $\hat{\Sigma} ^{(2)}$ &
$\hat{\Sigma}$ & $\hat{\Sigma}$ & Experiment \\
       &    &     & with fitting &      & with fitting & \\
\hline
$6s$ & 1425.0 & 1717.5 & 2306.9 & 2287.9 & 2285.7 &
2298.2 \tablenotemark[1] \\
$7s$ & 391.6 & 471.1 & 544.4 & 539.6 & 540.2 &
545.90(9) \tablenotemark[2] \\
$6p_{1/2}$ & 160.9 & 200.3 & 291.5 & 296.4 & 293.3 &
291.89(8) \tablenotemark[3] \\
$7p_{1/2}$ & 57.6 & 71.2 & 94.3 & 95.4 & 94.6 &
94.35 \tablenotemark[1] \\
\end{tabular}
\tablenotetext[1]{Ref. \cite{cshfs}.}
\tablenotetext[2]{Ref. \cite{gilbert83}.}
\tablenotetext[3]{Ref. \cite{rafac97}.}
\end{table}
\begin{table}
\caption{Percentage deviation from experiment of calculated
hyperfine structure constants in different approximations.}
\label{tab:hfsii}
\begin{tabular}{ldddd}
State &
\multicolumn{3}{c}{Percentage deviation} \\
 & $\hat{\Sigma}^{(2)}$ & $\hat{\Sigma}$ & $\hat{\Sigma}$ \\
 & with fitting    &          & with fitting \\
\hline
$6s$ & 0.4 & -0.4 & -0.5 \\
$7s$ & -0.3 & -1.2 & -1.0 \\
$6p_{1/2}$ & -0.1 & 1.5 & 0.5 \\
$7p_{1/2}$ & 0.05 & 1.1 & 0.3 \\
\end{tabular}
\end{table}
\begin{table}
\caption{Percentage deviation from experiment of calculated
$\sqrt{{\rm hfs}(s){\rm hfs}(p)}$ (we will denote this by $s-p$ in
the tables) in different approximations.}
\label{tab:hfsiii}
\begin{tabular}{lddd}
$\sqrt{{\rm hfs}(s){\rm hfs}(p)}$&\multicolumn{3}{c}{Percentage deviation}\\
 & $\hat{\Sigma} ^{(2)}$ & $\hat{\Sigma}$ & $\hat{\Sigma}$ \\
 & with fitting    &          & with fitting \\
\hline
$6s-6p$ & 0.1 & 0.5 & -0.02 \\
$6s-7p$ & 0.2 & -0.3 & -0.2 \\
$7s-6p$ & -0.2 & 0.2 & -0.3 \\
\end{tabular}
\end{table}
\begin{table}
\caption{Values for $E_{PNC}$ in different approximations;
units $10^{-11}iea_{B}(-Q_{W}/N)$.}
\label{tab:pncii}
\begin{tabular}{cddd}
 & $\hat{\Sigma}^{(2)}$ with fitting & $\hat{\Sigma}$ &
$\hat{\Sigma}$ with fitting \\
\hline
$E_{PNC}$ & 0.898 & 0.902 & 0.901 \\
\end{tabular}
\end{table}

\center
\widetext
\input psfig
\psfull

\begin{figure}[b]
\centerline{\psfig{file=hpscreense.eps, clip=}}
\caption{The electron self-energy operator with screening and hole-particle
interaction included.}
\label{fig:hpscreense}
\end{figure}

\begin{thebibliography}{20}

\bibitem{dzuba89}

V.A. Dzuba, V.V. Flambaum, and O.P. Sushkov,
Phys. Lett. A {\bf 141}, 147 (1989).

\bibitem{blundell90}

S.A. Blundell, W.R. Johnson, and J. Sapirstein,
Phys. Rev. Lett. {\bf 65}, 1411 (1990);
S.A. Blundell, J. Sapirstein, and W.R. Johnson,
Phys. Rev. D {\bf 45}, 1602 (1992).

\bibitem{wood97}

C.S. Wood {\it et al.}, Science {\bf 275}, 1759 (1997).

\bibitem{wieman99}

S.C. Bennett and C.E. Wieman, Phys. Rev. Lett. {\bf 82},
2484 (1999); {\bf 83}, 889 (1999).

\bibitem{derevianko00}

A. Derevianko, Phys. Rev. Lett. {\bf 85}, 1618 (2000).

\bibitem{sushkov01}

O.P. Sushkov, Phys. Rev. A {\bf 63}, 042504 (2001).

\bibitem{dzuba01}

V.A. Dzuba, C. Harabati, W.R. Johnson, and M.S. Safronova,
Phys. Rev. A {\bf 63}, 044103 (2001).

\bibitem{kozlov01}

M.G. Kozlov, S.G. Porsev, and I.I. Tupitsyn,
Phys. Rev. Lett. {\bf 86}, 3260 (2001).

\bibitem{marciano}

W.J. Marciano and A. Sirlin, Phys. Rev. D {\bf 27}, 552 (1983);
W.J. Marciano and J.L. Rosner, Phys. Rev. Lett. {\bf 65}, 2963 (1990).

\bibitem{lynnbed}

B.W. Lynn and P.G.H. Sandars, J. Phys. B. {\bf 27}, 1469 (1994);
I. Bednyakov, L. Labzowsky, G. Plunien, G. Soff, and V. Karasiev,
Phys. Rev. A {\bf 61}, 012103 (1999).

\bibitem{milstein}

A.I. Milstein and O.P. Sushkov, e-print hep-ph/0109257.

\bibitem{johnson01}

W.R. Johnson, I. Bednyakov, and G. Soff, submitted to Phys. Rev. Lett.;
e-print hep-ph/0110262.

\bibitem{derevianko}

A. Derevianko, submitted to Phys. Rev. A; e-print physics/0108033.

\bibitem{dzuba00}

V.A. Dzuba and V.V. Flambaum,
Phys. Rev. A {\bf 62}, 052101 (2000).

\bibitem{dzuba97}

V.A. Dzuba, V.V. Flambaum, and O.P. Sushkov,
Phys. Rev. A {\bf 56}, R4357 (1997).

\bibitem{groom00}

D.E. Groom {\it et al.},
Euro. Phys. J. C {\bf 15}, 1 (2000).

\bibitem{dzuba89energy}

V.A. Dzuba, V.V. Flambaum, and O.P. Sushkov,
Phys. Lett. A {\bf 140}, 493 (1989).

\bibitem{dzuba89e1hfs}

V.A. Dzuba, V.V. Flambaum, A.Ya. Kraftmakher, and O.P. Sushkov,
Phys. Lett. A {\bf 142}, 373 (1989).

\bibitem{dzuba87}

V.A. Dzuba, V.V. Flambaum, P.G. Silvestrov, and O.P. Sushkov,
J. Phys. B {\bf 20}, 1399 (1987).

\bibitem{dzuba}

V.A. Dzuba, in preparation.


\bibitem{fricke95}

G. Fricke {\it et al.},
At. Data and Nucl. Data Tables {\bf 60}, 177 (1995).


\bibitem{r_{np}}

A. Trzci\'{n}ska {\it et al.},
Phys. Rev. Lett. {\bf 87}, 082501 (2001).

\bibitem{berestetskii}

V.B. Berestetskii, E.M. Lifshitz, and L.P. Pitaevskii,
{\it Relativistic Quantum Theory}
(Pergamon Press, Oxford, 1982).

\bibitem{mohr}

P.J. Mohr and Y.-K. Kim, Phys. Rev. A {\bf 45}, 2727 (1992);
P.J. Mohr, Phys. Rev. A {\bf 46}, 4421 (1992).

\bibitem{fullerton76}

L.W. Fullerton and G.A. Rinker, Jr., Phys. Rev. A {\bf 13},
1283 (1976).

\bibitem{milstein83}

A.I. Milstein and V.M. Strakhovenko,
ZhETF {\bf 84}, 1247 (1983).


\bibitem{moore}

C.E. Moore, Natl. Stand. Ref. Data Ser.
(U.S., Natl. Bur. Stand.), {\bf 3} (1971).


\bibitem{rafac99}

R.J. Rafac, C.E. Tanner, A.E. Livingston, and H.G. Berry,
Phys. Rev. A {\bf 60}, 3648 (1999).

\bibitem{bouchiat84}

M.-A. Bouchiat, J. Guena, and L. Pottier,
J. Phys. (France) Lett. {\bf 45}, L523 (1984).

\bibitem{bennett99}

S.C. Bennett, J.L. Roberts, and C.E. Wieman,
Phys. Rev. A {\bf 59}, R19 (1999).


\bibitem{cshfs}

E. Arimondo, M. Inguscio, and P. Violino,
Rev. Mod. Phys. {\bf 49}, 31 (1977).

\bibitem{gilbert83}

S.L. Gilbert, R.N. Watts, and C.E. Wieman,
Phys. Rev. A {\bf 27}, 581 (1983).

\bibitem{rafac97}

R.J. Rafac and C.E. Tanner, Phys. Rev. A {\bf 56}, 1027 (1997).

\end{thebibliography}
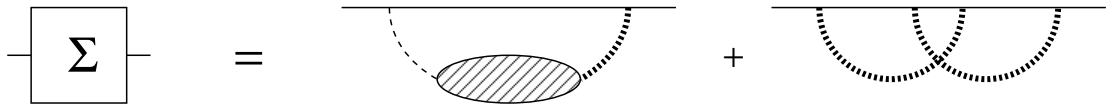
\end{document}